\documentclass[desactivate]{aa}  
\usepackage{graphicx}
\usepackage{txfonts}

\begin{document}

   \title{Lack of emission lines in the optical spectra of SAX J1808.4-3658 during reflaring of the 2019 outburst\thanks{ESO program ID: 2103.D-5054(A)}}


   \author{L. Asquini
          \inst{1}\fnmsep\inst{2}
          \and
          M. C. Baglio\inst{1}
          \and
          S. Campana\inst{1}
          \and
          P. D'Avanzo\inst{1}
          \and
          A. Miraval Zanon\inst{3}\fnmsep\inst{4}
          \and
          K. Alabarta\inst{5}
          \and
          D.M. Russell\inst{5}
          \and
          D.M. Bramich\inst{5}
          }

   \institute{INAF, Osservatorio Astronomico di Brera, Via E. Bianchi 46, I-23807 Merate (LC), Italy; \email{laura.asquini@inaf.it}
         \and
             Università dell’Insubria, Dipartimento di Scienza e Alta Tecnologia, Via Valleggio 11, I-22100, Como, Italy
         \and
          ASI - Agenzia Spaziale Italiana, Via del Politecnico snc, 00133, Rome, Italy
         \and 
          INAF - Osservatorio Astronomico di Roma, Via Frascati 33, I-00078, Monteporzio Catone (RM), Italy   
          \and
            Center for Astrophysics and Space Science (CASS), New York University Abu Dhabi, PO Box 129188, Abu Dhabi, UAE
             }

   \date{Received May 21, 2024; accepted March 16, 1997}

 
  \abstract
   {}
   {We present spectroscopy of the accreting X-ray binary and millisecond pulsar SAX J1808.4-3658. These observations are the first to be obtained during a reflaring phase. We collected spectroscopic data during the beginning of the reflaring of the 2019 outburst and compared them to previous datasets taken at different epochs, both of the same outburst and across the years. To this end, we also present spectra of the source taken during quiescence in 2007, one year before the next outburst.}
   {We made use of data taken by the Very Large Telescope (VLT) X-shooter spectrograph on August 31, 2019, three weeks after the outburst peak. For flux calibration, we used photometric data taken during the same night by the 1m telescopes from the Las Cumbres Observatory network that are located in Chile. We compare our spectra to the quiescent data taken by the VLT-FORS1 spectrograph in September, 2007. We inspected the spectral energy distribution by fitting our data with a multicolored accretion-disk model and sampled the posterior probability density function for the model parameters with a Markov chain Monte Carlo (MCMC) algorithm.}
   {We find the optical spectra of the 2019 outburst to be unusually featureless, with no emission lines present despite the high resolution of the instrument. 
   Fitting the UV-optical spectral energy distribution with a disk plus
irradiated star model results in a very large value for the inner disk radius of $\sim5130 \pm 240$ km,  which could suggest that the disk was emptied of material during the outburst, possibly accounting for the emission-less spectra. Alternatively, the absence of emission lines could be due to a significant contribution of the jet emission at optical wavelengths.}
   {}


   \maketitle
%

\section{Introduction}
    Low-mass X-ray binaries (LMXBs) are binary systems where a compact object, either a neutron star (NS) or a black hole, accretes matter from a low-mass ($<1 M_{\odot}$) companion star via Roche lobe overflow, forming an accretion disk. NS-LMXBs have been linked to millisecond pulsars (MSP) through the so-called recycling scenario \citep[e.g., ][]{srinivasan2010recycled}, according to which an old NS in the tight binary is re-accelerated through accretion of matter to the observed millisecond periods. The confirmation of this evolutionary channel came with the discovery of SAX J1808.4-3658 (SAX J1808 hereafter), which was first revealed by \emph{Beppo}SAX \citep{zand1998discovery} and then observed with the \emph{Rossi} X-ray Timing Explorer (RXTE; \citealt{wijnands1998millisecond}). This is an LMXB that was observed to exhibit a $401$ Hz coherent pulsation in an X-ray outburst, proving that it contains a fast-spinning NS.
    
    The source is located at a distance of 2.5-3.5 kpc. The NS orbits a semi-degenerate companion with a mass of $\sim$0.05--0.10 M$_{\odot}$ \citep{bildsten2001brown, wang2001optical,deloye2008optical}, and Doppler delays in the coherent timing data indicate an orbital period of 2.01 hours \citep{chakrabarty1998two}. SAX J1808 is a transient system; it stays in a quiescent state for $\sim$2-4 years and suddenly goes into outburst, with a rapid increase of its X-ray (and UV-optical) luminosity up to $2 \times 10^{36}$ erg s$^{-1}$ and ${\rm R}\sim 16.1$  mag \citep{zand2001first, roche1998sax}.
    
    In its quiescent state, this system has a very low X-ray luminosity ($5 \times 10^{31}$ erg s$^{-1}$, \citealt{campana2002xmm, hartman2008long}) and a faint optical magnitude of ${\rm R}\sim 21$ mag \citep{homer2001optical}, and it is associated with the optical-infrared variable V4584 Sagittarii. 
    The sinusoidal-shaped optical light curve, in quiescence, is thought to be produced by the tidally locked, irradiated companion star \citep{homer2001optical, deloye2008optical}. However, the observed X-ray flux is insufficient in accounting for the optical emission. The excess in the optical bands can be explained by invoking extra irradiation coming from an active rotational-powered radio pulsar \citep{campana2004indirect,burderi2003optical}. 
    
    The recurrent outbursts of this source, with ten registered events since its discovery, feature a pattern of increase of its X-ray luminosity by a factor of $10^5$ in a short period of time ($\sim$days). The increase in emission in the X-rays is preceded \citep{goodwin2020enhanced} by an increase of the UV-optical-infrared emission. After the outburst peak, the X-ray flux shows a slow decay lasting $\sim 0.5-1$ month and a subsequent reflaring for $\sim 10$ days. This last phase is thought to be linked to the so-called propeller scenario \citep{patruno2016reflares}, in which X-ray emission is influenced by the relation between the accretion rate and the magnetospheric radius. After this phase, the system goes back to its quiescent emission levels (e.g., \citealt{stella2000discovery}).

    In 2019, after the first optical brightening on July 30, SAX J1808 was confirmed to have begun a new outburst phase on August 6 (MJD 58701). The optical peak of this outburst was reached around August 10 (MJD 58705), and four days later (MJD 58709) the peak was reached in X-rays \citep{goodwin2020enhanced}. On August 24, observations in the X-rays by \cite{2019ATel130771B} determined that the source had entered its reflaring state, which was then observed in the optical by \cite{2019ATel131031B}. The authors reported four distinct flaring events on the following dates: August 30, September 2, September 4, and September 12. 

    In this paper, we present spectroscopy of the optical reflaring phase of SAX J1808, as well as its quiescent spectrum taken in 2007, which was one year before the 2008 outburst. In Sect. \ref{sec:2}, we describe the observations carried out with the Very Large Telescope (VLT) X-shooter and VLT-FORS1, while in Sect. \ref{sec:3} we describe the spectral properties of the two sets of spectra. In Sect. \ref{sec:4}, 
    we proceed to describe the spectral energy distribution (SED). In Sect. \ref{sec:5}, we discuss our findings, and we draw our conclusions in Sect. \ref{sec:6}. 

\section{Observations and data reduction}
\label{sec:2}

To perform our analysis, we used data from various instruments on different telescopes and covering multiple wavebands. A summary of the datasets used in this paper is provided in Table \ref{table:1}.

\subsection{2007 quiescence}
 We performed a spectroscopic observation of SAX J1808 during the quiescent state in September, 2007 from MJD 54345.9915 to 54346.1839, one year before the 2008 outburst. These were 22 spectra with 500 s exposure each, taken by the VLT-FORS1 instrument with a slit width of $1''$ and 300V grism (R$\sim$1650). 
 
\subsection{2019 outburst}
\begin{figure}
   \centering
   \includegraphics[width=0.5\textwidth]{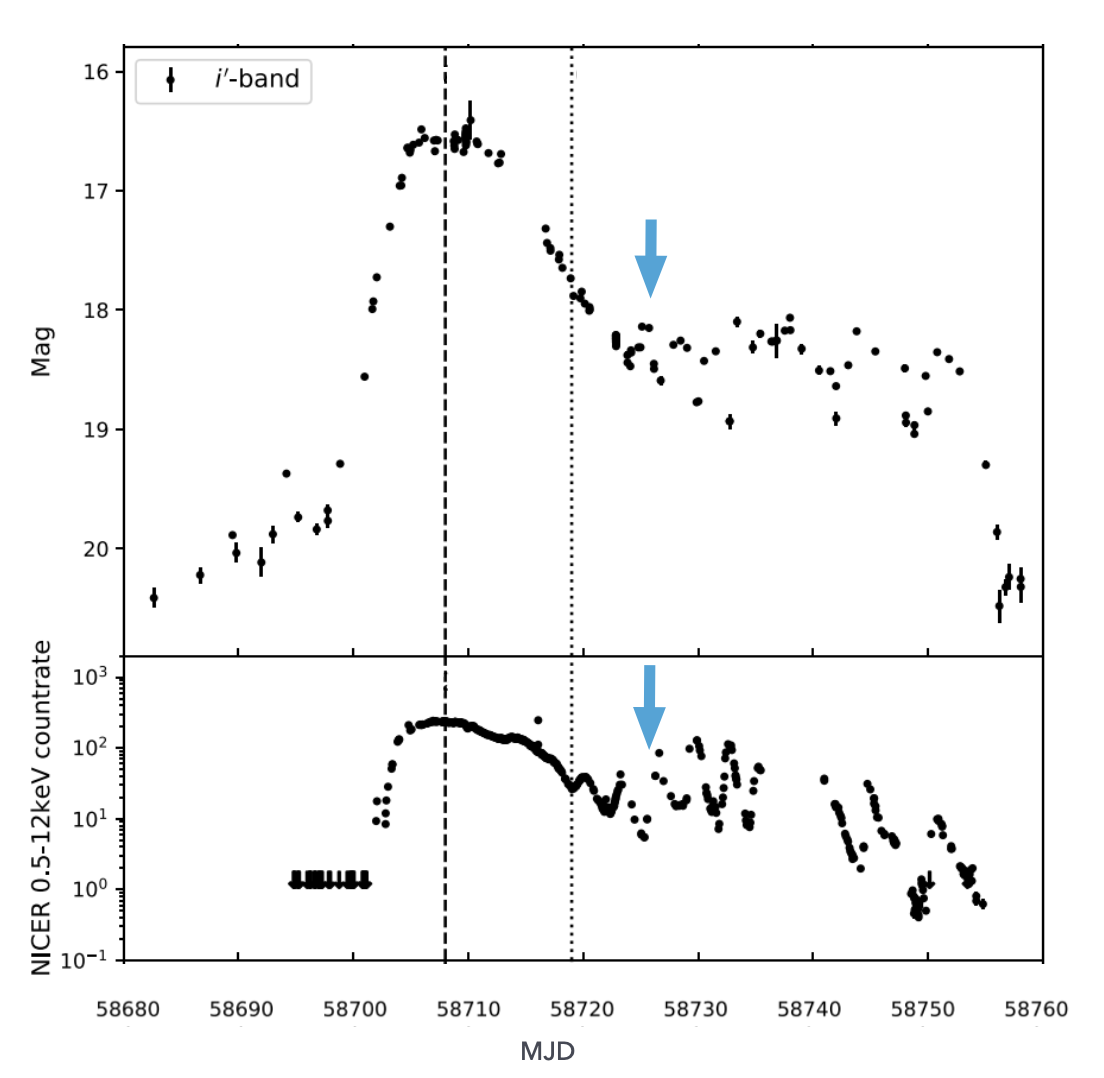}
   \caption{Optical \textit{i'-}band (top panel) and X-ray light curves (bottom panel) of the 2019 outburst, monitored, respectively, by LCO and NICER. Blue arrows point to MJD 58726.0093, when our X-Shooter spectra were acquired. The dashed line indicates the peak of the X-ray outburst, while the dotted line indicates the beginning of the X-ray reflaring state. Figure adapted from \cite{Baglio2020}.}
              \label{lightcurve}
    \end{figure}

Our monitoring of SAX J1808 started three weeks after the peak was reached in the optical, one week after the beginning of the X-ray reflaring, and one day after the first optical reflaring episode (Fig. \ref{lightcurve}). This is the first time that optical spectroscopy has been carried out during this phase. The observations were carried out by the X-shooter spectrograph \citep{vernet2011x} at the VLT on August 31, 2019 (MJD 58726.0093) from 00:13 to 02:41 UTC (MJD 58726.11205), covering more than one orbital period of the binary. The instrument has three different arms that can observe in the ultraviolet (UV, R$\sim$5400), optical (VIS, R$\sim$8900), and near-infrared (NIR; R$\sim$5600) domains, covering between 300 and 2480 nm. We used a $0.9''$ slit width for UV and VIS arms, and a $1''$ slit width for the NIR.  We obtained 19 spectra in STARE mode for each arm, with 300 seconds of exposure each. We reduced the spectra using the X-Shooter pipeline \citep{modigliani2010x}, following the usual procedures for bias and flat-field correction. We also collected UV spectra on August 28, 2019 (MJD  58723.9076) with the Space Telescope Imaging Spectrograph (STIS), operating in the 165-310 nm UV band, on-board the \textit{Hubble} Space Telescope (HST; STIS, GO/DD-15987, PI Miraval Zanon, \citealt{ambrosino2021optical}). As the authors describe, they performed observations in TIME-TAG mode with 125 $\mu$s time resolution for about 2.2 ks with the NUV-MAMA detector. They used the G230L grating and equipped it with a $52\times0.2$ arcsec slit, thus ensuring a spectral resolution of $\sim500$ over the nominal range (500-1010). They estimated the background signal by selecting and averaging photons  outside the source region in the instrument slit channels, normalizing the result over the total number of channels. 
All spectroscopic data were analyzed using the MOLLY package\footnote{http://deneb.astro.warwick.ac.uk/phsaap/software/}.

On the same date that the X-Shooter spectra were taken, optical imaging was acquired with the 1m telescopes from the Las Cumbres Observatory network that are located in Cerro Tololo (Chile) in the z, i', V, R, and B bands, with exposure times of 200s in z and 100s in all the remaining filters. All observations were taken between MJD 58726.1509 and MJD 58726.1602. The data were processed through the newly developed X-Ray Binary New Early Warning System (XB-NEWS)
pipeline \citep{Russell2019}, which performs multi-aperture photometry (MAP), solves for zero-point
calibrations between epochs \citep{bramich2012systematic}, and flux-calibrates the photometry using the ATLAS-REFCAT2 catalog (which includes the Pan-STARRS DR1,\footnote{\url{https://panstarrs.stsci.edu}} APASS,\footnote{\url{https://www.aavso.org/apass}} and other catalogs; \citealt{Tonry2012}). In contrast to the B, V, i', and z bands, the R band is calibrated indirectly via comparison with
predicted R-band standard magnitudes. The predicted R-band
standard magnitudes are computed for ATLAS-REFCAT2
stars with Pan-STARRS1 $g_{P1}$ and $r_{P1}$ standard magnitudes
using the transformations provided in \cite{Tonry2012}. The
pipeline then produces a calibrated light curve for the target
with near real-time measurements (for more details, see
\citealt{Russell2019}). We note that an almost simultaneous observation in the u' filter (central wavelength: $3540 \AA$) also exists. However, we exclude it from this analysis due to the absence of u'-band standard stars in the field.

In addition to the optical data, we retrieved UV photometric data from the \textit{Swift}-UVOT archive ($uvw1$, $uvm2$, $uvw2$ filters). The closest data to our dataset were obtained on MJD 58725.0738 (2019-08-30 01:46:14 UTC), with an exposure time of 518.89 s. We note that according to the B-band LCO optical photometry, only minimal ($<0.05$ mag) variations in the fluxes between August 29 and September 2, are observed, which justifies using the UVOT data for our study. To extract the magnitudes, we used the {\tt uvotsource} HEASOFT routine,
defining a circular aperture centered on
the source with a radius of $3''$ as the extraction region and a circular
aperture (away from the source) with a radius of $10''$ as background. This relatively small radius has been used to avoid nearby sources.
After correcting for reddening (using $A_V=0.51\pm 0.04$ mag as reported in \citealt{Baglio2020} and \citealt{goodwin2020enhanced} and using the relations of \citealt{cardelli89}), we derived the flux densities for unit frequencies in all bands.

To compute the high-energy luminosity \textit{$L_{\rm irr}$} irradiating the companion star, we considered the observations collected on August 31, 2019 (MJD 58726.2718, with an exposure of 1542.0 s) by the Neutron Star Interior Composition Explorer \citep[NICER;][]{gendreau2016neutron}. This dataset is the closest one to our X-shooter observations and is identified by the ObsID 2584011501 in the NICER archive database. 
We analyzed it using the software HEASOFT version 6.32.1 and NICERDAS version 11a. The CALDB (archival calibration data) version used was 20221001. We applied standard filtering and cleaning criteria. We included the data when the dark Earth limb angle was $>15\deg$, the pointing offset was $< 54''$, the bright Earth limb angle was $>30\deg$, and the International Space Station (ISS) was outside the South Atlantic Anomaly. We removed data from detectors $14$ and $34$ since they show episodes of increased electronic noise. We extracted a background-subtracted energy spectrum using the {\tt nibackgen3c50} model \citep{Remillard22}. We fitted the energy spectra of SAX J1808 using XSPEC \citep[V. 12.10.1;][]{arnaud96}. We only considered the energy band $1.0-10.0 \, \rm keV$ and included a Gaussian absorption line at 0.76 keV to model the instrumental residuals below $\sim3$ keV that are typical for X-ray missions and Si-based detectors \citep[e.g.,][]{Ludlam2018, Miller18}. We then rebinned the spectra to have at least 25 counts per bin. We fit the spectrum with a combination of an absorbed power-law model and a blackbody. As a result of the fit, we obtain a column density of $N_{H} = (1.1\pm0.1)\times 10^{21} \rm \ cm^{-2}$, a power-law index of 2.02 $\pm$ 0.03, and a blackbody temperature of $(0.41\pm0.01) \, \rm keV$ and radius is $\sim 3$ km (considering a distance of 2.5 kpc), which is consistent with an origin in the neutron star. The fit is acceptable with $\chi^{2} = 162.7$ and $127$ degrees of freedom. The unabsorbed X-ray luminosity extrapolated from the $0.1-10 \ \rm keV$ range is $ 2.2 \times 10^{35}\ \rm erg\ s^{-1}$. 

\begin{table*}
\caption{Datasets used in this paper.}             
\label{table:1}     
\centering                          
\begin{tabular}{l c c c c r}        
\hline\hline 
Instrument & Telescope & Type & Band & Start of observations [MJD] & Exposure time [s] \\    
\hline                       
   X-shooter & VLT & Spectroscopic & NIR, VIS, UVB & 58726.0093 & 19 $\times$ 300\\
   STIS & HST &  Spectroscopic   & UV &        58723.9076 & 2200/125$\mu$s \\
   fa15 & 1m0-05 & Photometric & $z, i', R, V, B$ & 58726.1509 & 200 ($z$), 100 \\
   UVOT & \textit{Swift} & Photometric    & UV & 58725.0738 & 518.89\\
   XTI & NICER & Spectroscopic & X-ray & 58726.2718 & 1542.0 \\
\hline 
   FORS1 & VLT & Spectroscopic    & NIR-UVB & 54345.9915 & 22 $\times$ 500 \\
\hline                                   
\end{tabular}
\tablefoot{The top panel shows data used for the analysis of the 2019 outburst, while the bottom panel displays data from the observations in quiescence in September, 2007.}
\end{table*}

\section{Spectral Properties}
\label{sec:3}
\subsection{2007 quiescence}
\begin{figure*}
   \centering
   \includegraphics[width=0.7\textwidth]{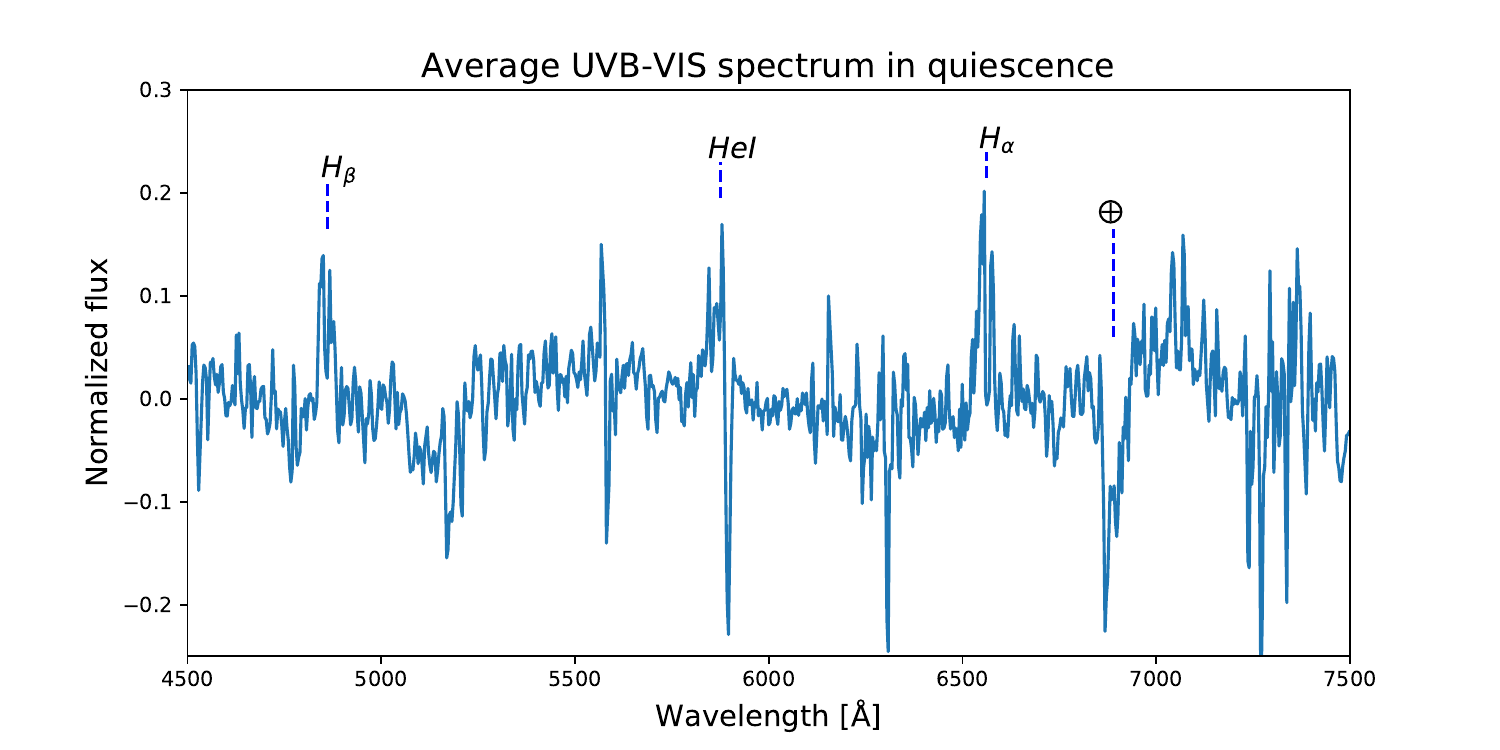}
   \caption{Average optical spectrum of SAX J1808 in quiescence (MJD 54345.9915). The spectrum presents emission lines from H$_{\alpha}$, H$_{\beta,}$ and HeI. The cross at $\lambda6900$\AA\ indicates a telluric line.}
    \label{spec_quiesc}
    \end{figure*}

\begin{figure}
   \centering
   \includegraphics[width=0.5\textwidth]{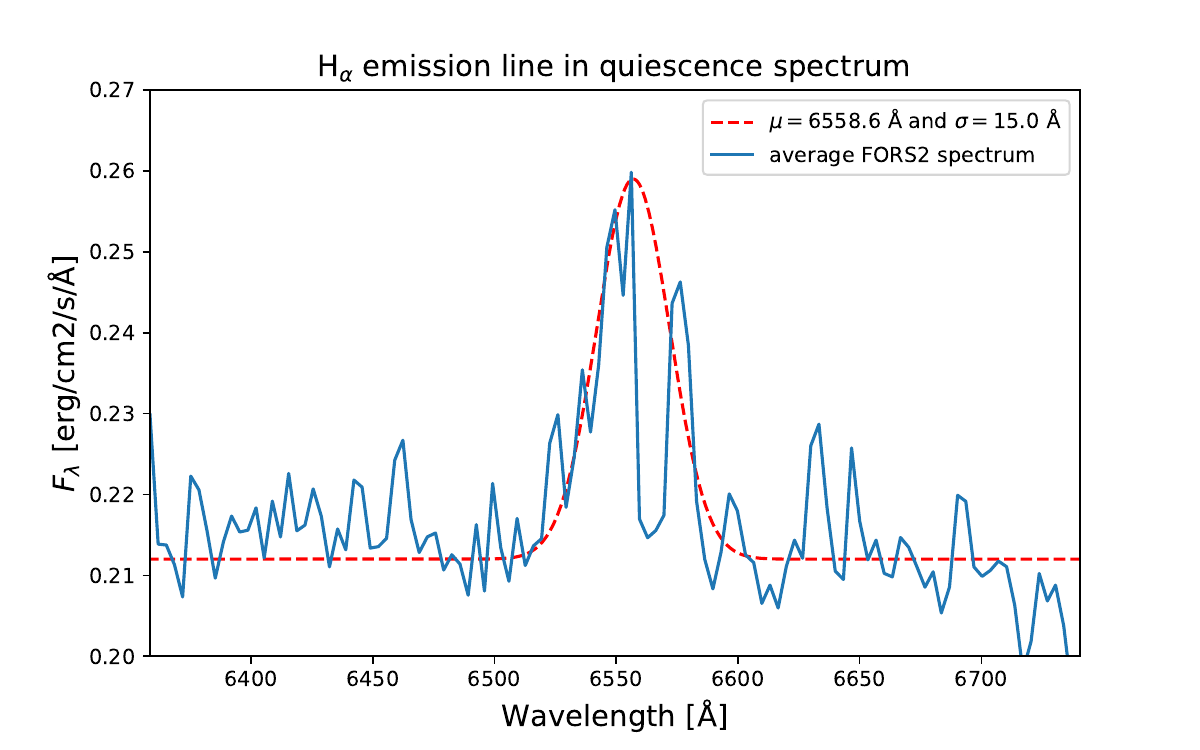}
   \caption{Broad H$_{\alpha}$ emission line during quiescence in September, 2007 (MJD 54345.9915). The blue solid line shows the VLT-FORS1 spectrum,  while the dashed red line shows a Gaussian fit over the line, excluding the core. }
    \label{halphaqui}
    \end{figure}

Each of the acquired spectra displayed prominent emission features from  H$_{\alpha}$, H$_{\beta,}$ and HeI at $\lambda$5875\AA\ (Fig. \ref{spec_quiesc}), adjacent to the interstellar NaD absorption line.
The emission lines appear to be broadened and show a double-horned profile, indicating their origin to be in an accretion disk. To better evaluate the width of the line wings (which are a proxy of the fastest, innermost part of the disk), we fit all of these lines with a single Gaussian component, neglecting the points in the core. The emission lines present a full width at half maximum of around $\sim$35$\pm$5.4\AA, corresponding to velocities of roughly 1600$\pm$247 km s$^{-1}$,  which is consistent with inner-disk velocities.  At these wavelengths and with our setup, FORS1 is capable of resolving velocities down to 700 km s$^{-1}$, making our data well above the resolving power of the instrument. In Fig. \ref{halphaqui}, we show the H$_{\alpha}$ emission, which is the line with the highest signal-to-noise ratio.

\subsection{2019 outburst}
\begin{figure}
   \centering
   \includegraphics[width=0.5\textwidth]{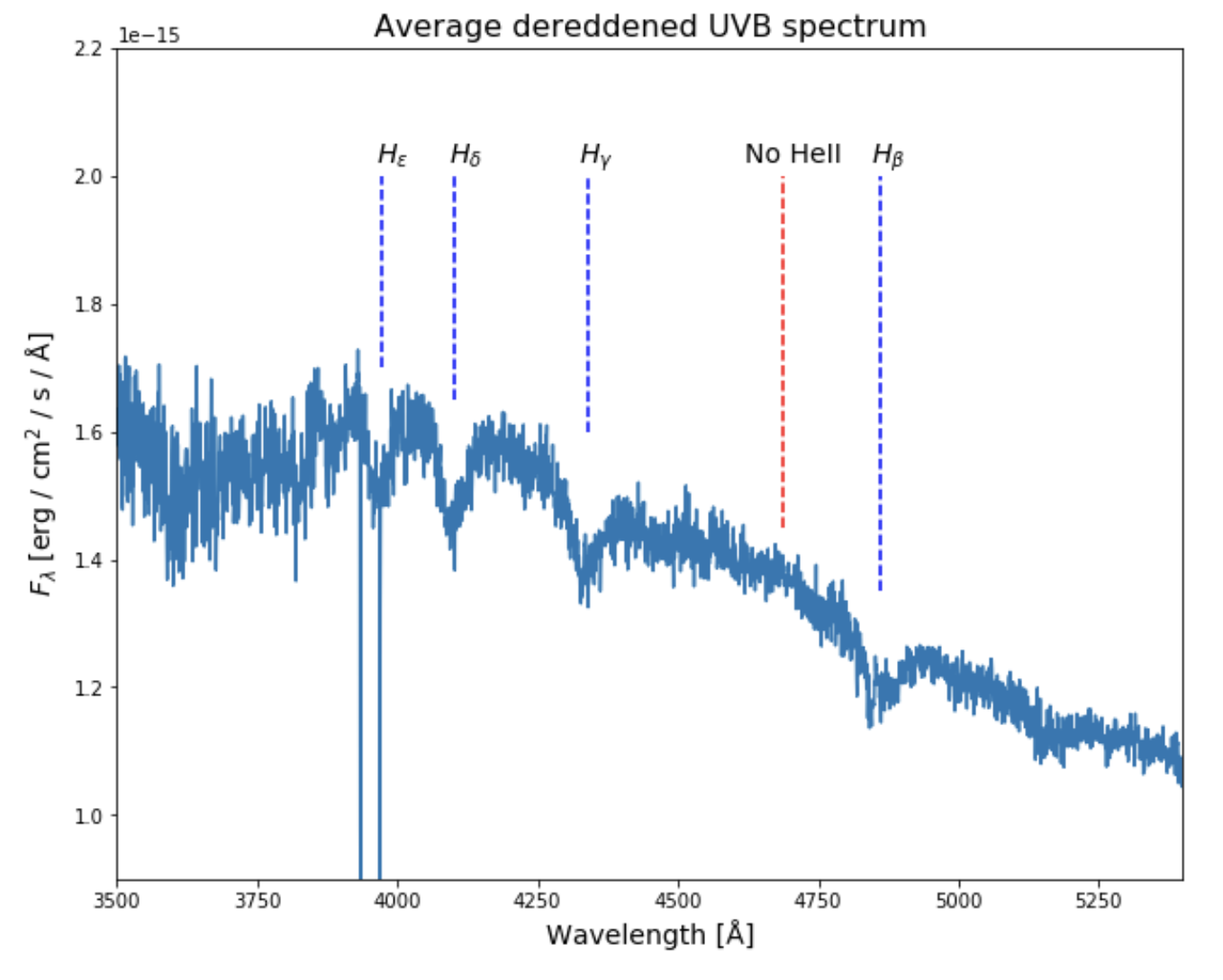}
   \caption{Average X-shooter UVB spectrum of SAX J1808, acquired on MJD 58726.0093 and corrected for Galactic absorption. The UVB arm is scaled to match the VIS flux. There are obvious absorption lines from the Balmer series, but no characteristic HeII emission line or Bowen complex ($\lambda$4630-4660\AA).}
              \label{avg_uvb}
    \end{figure}

\begin{figure}
   \centering
   \includegraphics[width=0.5\textwidth]{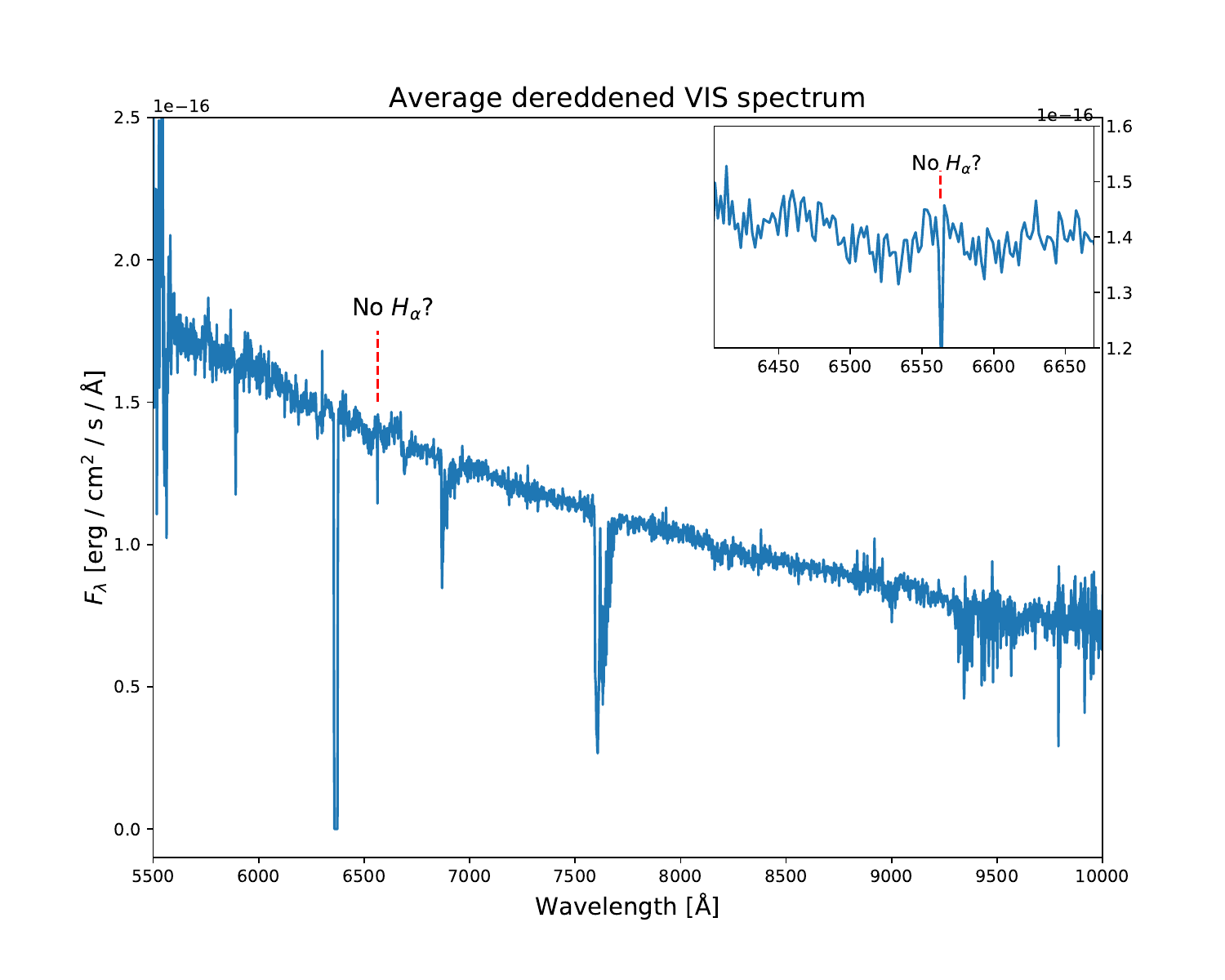}
   \caption{Average X-shooter optical spectrum of SAX J1808, from observations on MJD 58726.0093. The spectrum is substantially featureless, except for telluric absorptions throughout. The top right inset shows a zoomed-in view of wavelengths around the missing H$_{\alpha}$. Only a very thin dip is present at $\lambda$6563\AA.}
              \label{visspec}
    \end{figure}

In Fig. \ref{avg_uvb}, we plot the average spectrum of SAX J1808 in the X-shooter UVB arm.  We corrected for Galactic reddening following \cite{schlafly2011measuring} within MOLLY with an $A_V=0.51 \pm 0.04$ mag. The UVB spectrum is dominated by broad Balmer lines in absorption, with central emission reversal in the H$_{\beta}$ line and the interstellar NaD absorption still present. We detected no emission lines from HeII at $\lambda$4686\AA, \ HeI at $\lambda$5875\AA,\, or from the Bowen complex at $\lambda$4630–4660\AA. Surprisingly, the H$_{\alpha}$ line was also missing from the visible spectra (Fig. \ref{visspec}), except for what could perhaps be interpreted as a weirdly thin absorption line.

\section{Spectral energy distribution fitting}
\label{sec:4}
We performed a fit of the UV-optical spectral energy distribution (SED), which allowed us to better characterize the overall broadband spectrum of the source obtained at multiple wavelengths in the same time window. By analyzing the SED over several orders of magnitude in frequency, we were able to identify the different physical processes that give origin to the observed emissions, and we thus had the chance to disentangle the contribution of the various components of the system (i.e., neutron star, disk, companion, hot spot, and so on).

\begin{figure}
   \centering
   \includegraphics[width=0.5\textwidth]{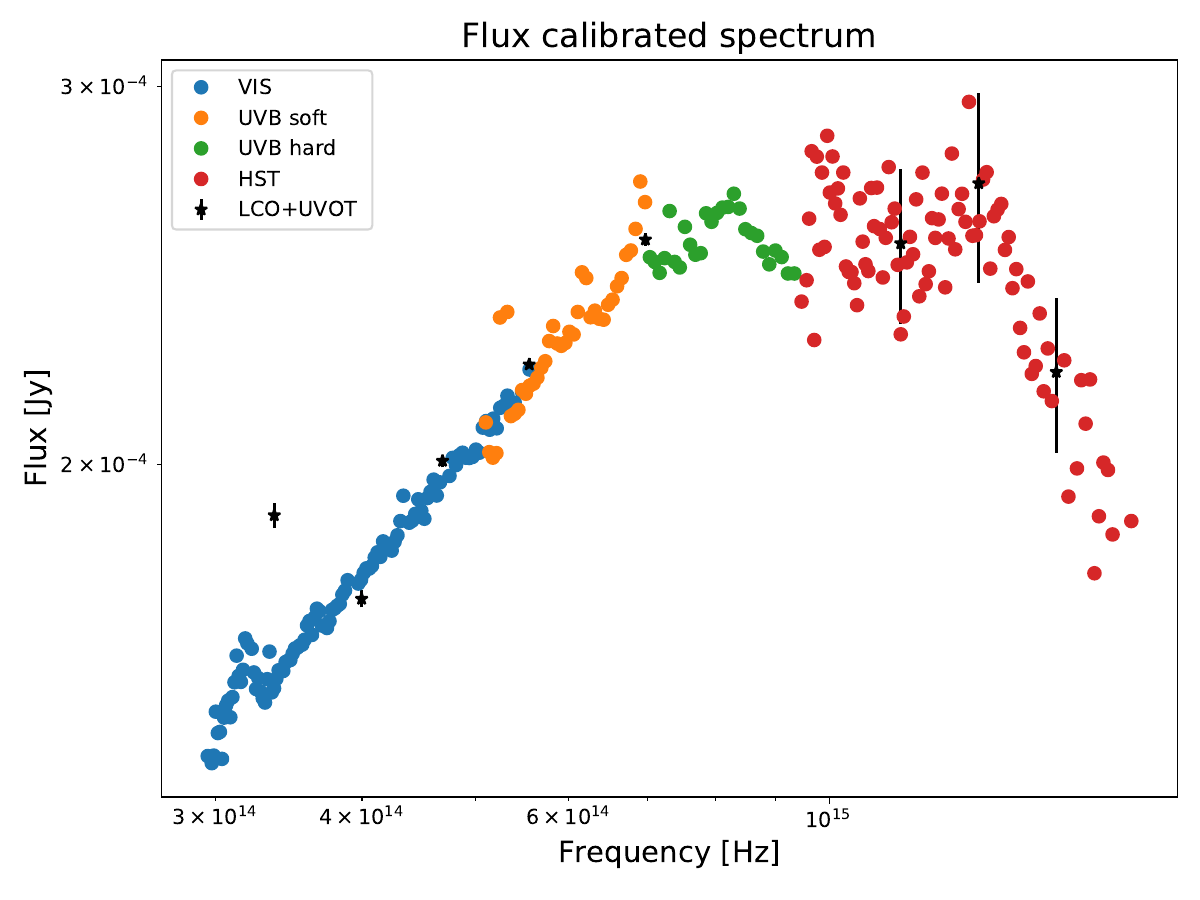}
   \caption{Flux-calibrated UVB-VIS spectra (observations started on MJD 58726.0093),  with fixed intrinsic scattering in UVB and VIS data and SED approximated as a superposition of power laws.}
              \label{scatpl}
    \end{figure}

Since the UVB-VIS spectra were not taken at the parallactic angle due to the field of SAX J1808 being particularly crowded, our spectra were affected by a further reddening effect that becomes important at blue wavelengths. In order to perform the flux calibration, we approximated the SED curve as a superposition of power laws, with their slopes and intercepts extrapolated from the LCO photometry with the i', R, V, and B points (in the log-log plot). We divided our data into segments using the frequency of the LCO data points to define their extremes, and we computed a power-law fit to each segment. By fixing the distance of each point from the resulting power-law fit, we fixed the intrinsic scattering of the data and reproduced it on the set of power laws dictated by the photometry, thus obtaining the plot (Fig. \ref{scatpl}). For HST data, we simply rescaled the data by a factor of 1.25 to match the UVOT photometric data. 

To perform the fit to the SED, we used a multicolored accretion-disk model to describe the optical - UV fluxes (following \citealt{Baglio2023}). For the multicolored accretion-disk model (Eqs.\,10--15 by \citealt{chakrabarty98}), we allowed the inner radius of the accretion disk ($r_{\rm in}$), the X-ray albedo of the disk (expect to be $\sim 0.95$ \citealt{chakrabarty98}), and the mass transfer rate from the companion to vary (expected to be $\sim 10^{-10}\,M_{\odot}\rm \, yr^{-1}$, predicted for an X-ray binary system with an orbital period of $\sim 2$ hr hosting a $0.1 M_{\odot}$ star and transferring mass through loss of angular momentum; \citealt{verbunt93}).
The distance to SAX J1808 ($D$), the binary separation ($a$), and the irradiation luminosity ($L_{\rm irr}$) were fixed at known (or reasonable) values: $D=2500$\,pc \citep{cornelisse2001first} and $a=[G(M_{\rm NS}+M_{\rm C})P_{\rm orb}^2/(4\pi)^2]^{1/3}$, where $M_{\rm NS}=1.7M_{\odot}$ is the NS mass, $M_{\rm C}=0.1\,M_{\odot}$ is the mass of the companion star, $P_{\rm orb}=2.01$\,hr is the binary orbital period, and $G$ is the gravitational constant. 
$L_{\rm irr}$ was fixed at $2\times10^{35} \, \rm\ erg\ s^{-1}$, which was calculated from the NICER observation taken on August 31, 2019.

We did not include the X-shooter VIS spectrum points below the i'-band frequency in the fit, as it has been shown by \citet{Baglio2020} that a non-negligible contribution from a compact jet might be expected at frequencies below the i' band. For similar reasons, we excluded the entire NIR spectrum. 
    
\begin{figure}
   \centering
   \includegraphics[width=0.5\textwidth]{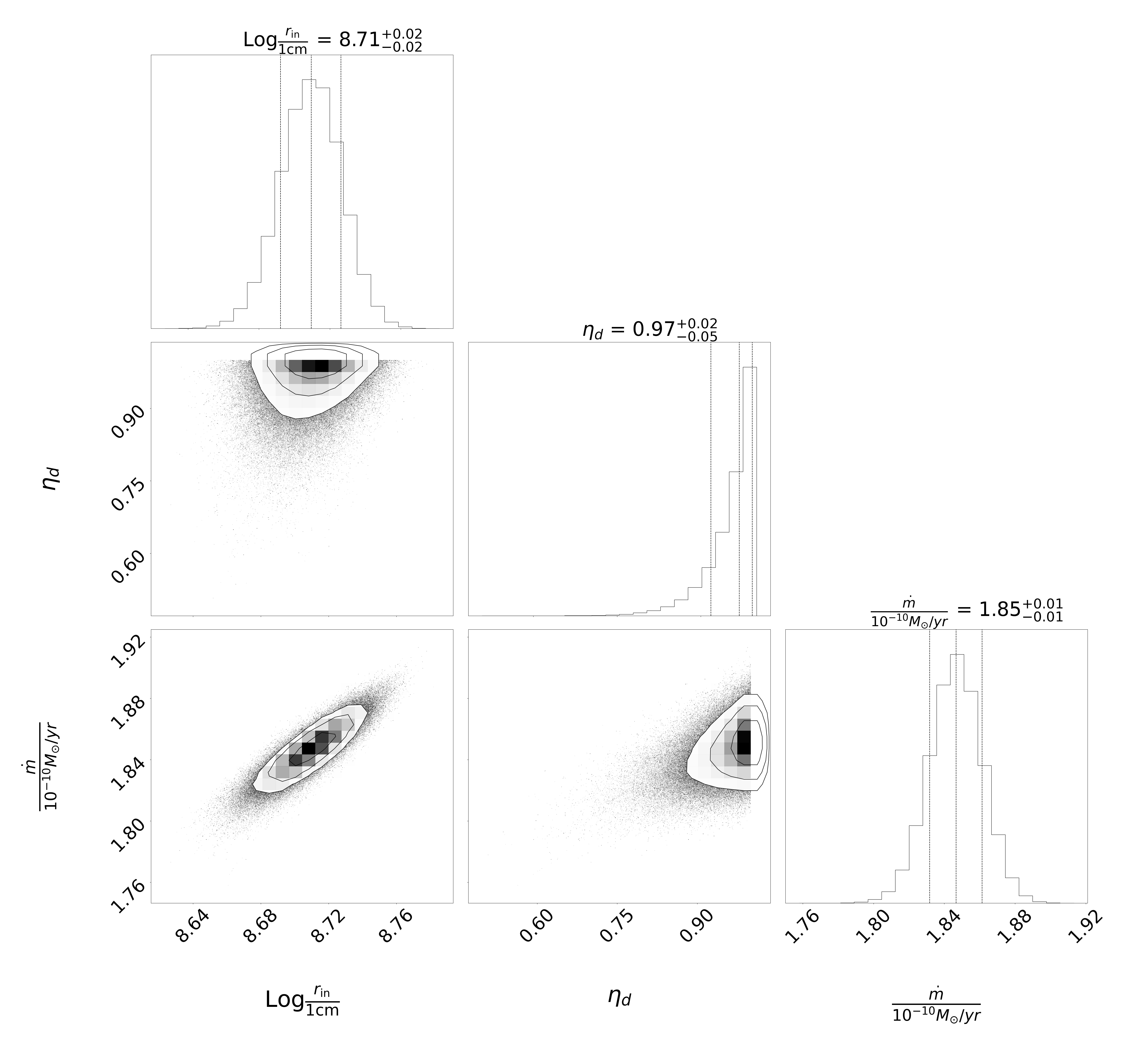}
   \caption{Corner plot of MCMC sampling. The plots off the diagonal show the 2D posterior distribution of the free parameters, namely $\rm Log\, \frac{R_{\rm in}}{1\rm cm}$ (left), $\eta_{D}$ (centre), and $\frac{\dot{m}}{10^{-10}}$ (right). The top panels of all three columns show the marginal posterior distribution used to estimate the free parameters.}
    \label{corner}
    \end{figure}

\begin{figure}
   \centering
   \includegraphics[width=0.5\textwidth]{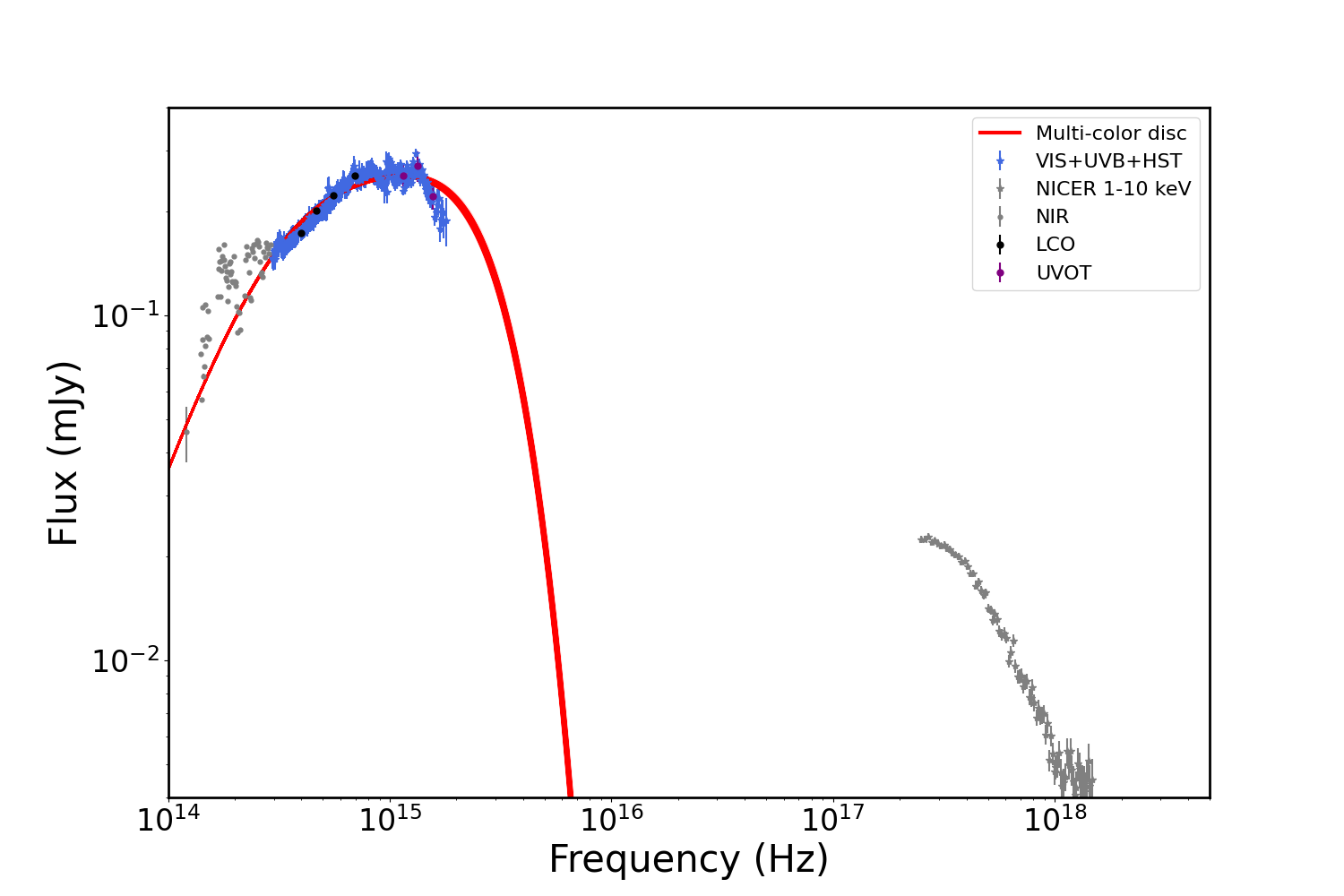}
   \caption{Spectral Energy distribution for SAX J1808 at the end of the 2019 outburst. VIS-UVB data were taken with X-shooter (MJD 58726.0093), UV data with {\it HST} (MJD 58723.9076), and X-ray data with {\it NICER} (MJD 58726.2718). Calibration points were retrieved with LCO (MJD 58726.1509) and \textit{Swift}-UVOT (MJD 58725.0738). The red solid line shows our MCMC fit to UV-optical data.}
              \label{sed}
    \end{figure}

We performed a Markov chain Monte Carlo (MCMC) sampling of the posterior probability density function for the parameter space of the model (Table \ref{table:2}). 
We have three free parameters in the fit, which are the inner disk radius $\rm Log\, \frac{R_{\rm in}}{1\rm cm}$, the companion star albedo $\eta_{D,}$ and the accretion rate $\frac{\dot{m}}{10^{-10}}$. For the radius, we used a uniform prior in the reasonable interval $\rm Log\, \frac{R_{\rm in}}{1\rm cm}\sim \mathcal{U}(2, Log\, \frac{R_{\rm out}}{1\rm cm})$, where $R_{\rm out}\sim 0.38\, a$ \citep{powell2007mass}. For the albedo, we used $\eta_{D} \sim \mathcal{U}(0.5, 1)$ as an
interval, and for the accretion rate the interval was  $\frac{\dot{m}}{10^{-10}} \sim \mathcal{U}(0.5, 2)$; this allowed, for all three variables, an ample range of the parameter space. This is of great importance to minimize the probability that the actual source parameters fall outside the region sampled by the chosen prior. As the corner plot in Fig. \ref{corner} shows, all three parameters are well constrained by the fit. Each of them has been estimated as the median of the marginal posterior distribution, with $1\sigma$ credible intervals coming from the 16th--84th percentiles of the posterior distribution. The results are $\rm Log\, \frac{R_{\rm in}}{1\rm cm}=8.71\pm 0.02$, translating to an inner disk radius of $\sim5.1 \times 10^{8}$ cm. The albedo results in $0.97^{+0.02}_{-0.05}$ , while the accretion rate converges at $1.85^{+0.01}_{-0.02}\times 10^{-10}$$ M_{\odot} \rm yr^{-1}$. Both values are in agreement with expectations. The fit to the data is shown in Fig. \ref{sed}.

\begin{table}
\caption{Parameters for the MCMC simulations.}             
\label{table:2}      
\centering                          
\begin{tabular}{l c r}        
\hline\hline 
Parameter & Posterior & Prior \\    
\hline                        
   $\rm Log\, \frac{R_{\rm in}}{1\rm cm}$ & $8.71^{+0.02}_{-0.02}$ & $\rm Log\, \frac{R_{\rm in}}{1\rm cm}\sim \mathcal{U}(2, Log\,r_{\rm out})$\\
   $\eta_{D}$ & $0.97^{+0.02}_{-0.05}$ & $\eta_{D} \sim \mathcal{U}(0.5, 1)$ \\
   $\frac{\dot{m}}{10^{-10}} \rm [ M_{\odot} \rm yr^{-1}\rm ]$ & $1.85^{+0.01}_{-0.02}$ & $\frac{\dot{m}}{10^{-10}} \sim \mathcal{U}(0.5, 2)$\\
\hline                                  
\end{tabular}
\end{table}
    
\section{Discussion}
\label{sec:5}
The spectra we analyzed were acquired at the beginning stages of the optical reflaring, three weeks after the optical peak of the main outburst. They show a remarkable absence of emission lines while simultaneously exhibiting very broad and deep absorption lines from the Balmer series, suggesting an origin from the fast-rotating disk. Previous observations by \cite{elebert2009optical} and \cite{cornelisse2009phase} clearly detect the typical HeII line and the Bowen blend during the peak of the 2008 outburst, which nonetheless showed similar absorption characteristics to the ones we found. The early time spectrum ($\sim$ 4 days before the optical peak) by \cite{goodwin2020enhanced} for the 2019 outburst, however, highlights that the only detected emission line at this epoch, HeII, was exceptionally weak. 

Besides the absence of Helium, the complete absence of H$_{\alpha}$ in our reflaring-phase spectrum is perhaps even more puzzling. Although in the literature (and also in \citealt{goodwin2020enhanced}) a weak H$_{\alpha}$ (either in absorption or emission) has been explained as a consequence of the low inclination of the system, this geometrical explanation would be in contradiction with the detection of the H$_{\alpha}$ line during quiescence, since the binary does not change its viewing angle. One might argue that the origin of this specific line is not in the disk, but in another location (e.g., the shock caused by the relativistic pulsar wind - if active - or the companion star); however, this would be in contradiction with the fact that the line is evidently broad (see Fig. \ref{halphaqui}, which shows a high dispersion despite the low resolution of the instrument).
We propose two possible explanations for the absence of the H$_{\alpha}$ line in our reflaring-phase spectra. 

\subsection{Emptied disk}
The first hypothesis to explain the lack of emission lines is that the inner region of the disk that typically produces the emission at optical wavelengths might be depleted of material because, at the end of the main outburst, the disk results emptied out. This interpretation is reinforced by the MCMC fit performed on the X-shooter-HST spectra, which converged to an inner disk radius of $\sim5130 \pm 240$ km,
painting the picture of a very thinned-out disk. However, we note that this result could be biased by the fact that our model only fits the portion of the accretion disk where the optical-UV is relevant.  However, the region where the inner disk radius is typically located would emit predominantly in the far-UV. Unfortunately, our dataset does not extend that far in the spectrum. Also, Fig. \ref{sed} shows that the fit fails to describe the emission at frequencies higher than $\sim 1.34\times 10^{15}\, \rm Hz$ (i.e., the UVOT $m2$ filter), where the spectrum shows a cutoff. This could indicate a missing puzzle piece. The apparent cutoff of the highest frequencies in our HST data could be explained by an irradiated inner disk, which would be better described by a different blackbody temperature than that of the optical disk; yet,
the lack of data in the far-UV frequency range makes it impossible to prove this.

\subsection{Jet contribution}
The second explanation relies on the fact that these data were acquired just one day after the first optical reflaring event was reported, two days before the next one, and a week after the beginning of the X-ray reflaring \citep{2019ATel130771B,2019ATel131031B}. The large value of the inner disk radius could be consistent with a propeller effect, whose onset would be triggered by the diminished accretion and the expulsion of the innermost portion of the disk. This, in turn, could be consistent with the large value of the inner-disk radius returned by our MCMC simulation. In their analysis of the 2019 outburst,
\citet{Baglio2020} also took into account data acquired from later epochs of the reflaring phase. They find that at these stages (see Epochs 7 and 8 in their paper), the contribution of the jet emission is non-negligible in the R, i', and z bands. Indeed, we see a clear excess for the LCO z calibration point in Fig. \ref{scatpl}, which we chose to neglect in order not to bias the MCMC fit, possibly due to a significant contribution of the jet (that is not included in our model function). Following the model for the jet contribution shown in Fig. 11 of \citet{Baglio2020} and their conclusions (at the time of our dataset) we likely meet the condition of a ``moderate'' jet, which contributes to $\sim 50\%$ of the flux in the R and i' bands and then increases with wavelength; our red-end spectrum would therefore have features that are inevitably camouflaged by the presence of the jet. In particular, the H$_{\alpha}$ emission line might be diluted by the additional red component (but not the $H_{\beta}$, which is found at higher frequencies). This hypothesis is further supported by the fact that the Balmer decrement H$_{\alpha}$/H$_{\beta}$ (after de-reddening) is typically constant at $\sim 3$ (from case B recombination theory; e.g. \citealt{Osterbrock2006}); the fact that we detect $H_{\beta}$ strongly in absorption, while there is no H$_{\alpha}$, suggests that the disk emission is diluted by the jet emission at the wavelength of the H$_{\alpha}$. This is consistent with most SEDs in the reflaring state published in \citet{Baglio2020}, and also with their color-magnitude diagram, which clearly shows the additional red component with respect to the accretion-disk blackbody also during this part of the outburst.

\section{Conclusions}
\label{sec:6}
 We present the first UV-optical spectra gathered during the beginning of the optical reflaring phase in the 2019 outburst of the transient low-mass X-ray binary SAX J1808.4-3658. The spectra were obtained on August 31, 2019, 21 days after the main outburst peak at optical wavelengths, and they show broadened absorption lines corresponding to the Balmer series, with a peculiar absence of emission-line features. Earlier observations by \cite{goodwin2020enhanced} before the outburst peak found a similar lack of emission lines (a weak HeII emission, no Bowen complex, and a broad but shallow H$_{\alpha}$ absorption). Besides the lack of emission, the comparison between the two epochs shows that the overall spectral properties of the optical reflaring stage look very similar to those that are observed during the main outburst. 

 While the weakness of the emission lines before the outburst peak can be explained by a disk that is not yet filled up, the explanation for our reflaring-phase spectra is possibly more elusive.  Given the presence of these emission-line features in quiescence (e.g., the 2007 spectra presented in this work), it is hard to attribute their absence to a low inclination of the system, so it appears more likely to be due to the conditions in the disk at this epoch. One possibility is that the disk is emptied because of the outburst itself. This idea is supported by the disk parameters we retrieved by fitting the calibrated optical-UV-VIS spectra with an MCMC algorithm; however, this conclusion might be biased by the lack of data at higher frequencies, which is where the inner disk usually lies. The fit may have converged to an inner disk value of $\sim5130 \pm 240$ km, which is indeed consistent with a hot, thinned-out disk. Another possibility is that the jet contribution, which is shown to be present and non-negligible in the reflaring phase \citep{Baglio2020}, is actively diluting the spectral features at wavelengths in the z, i', and $R$ bands. In addition to possibly explaining the lack of emission, the large value for the inner disk radius we obtained, associated with the fact that the data were taken during the reflaring phase, could point toward a propeller mechanism in action, as suggested earlier to explain the reflares in this system \citep{patruno2016reflares}.
To further explore these scenarios, it would be useful to apply an observation strategy on AMXP outbursts that replicate the same fortuitous conditions as these; that is, very early-time and very late-time observations, possibly with an additional dataset during the peak.  

\begin{acknowledgements}
MCB acknowledges support from the INAF-Astrofit fellowship. 
This material is based upon work supported by Tamkeen under the NYU
Abu Dhabi Research Institute grant CASS.
\end{acknowledgements}

%
%

\bibliographystyle{aa}
\bibliography{biblio}

\end{document}